\documentclass[%
11pt,
reprint,
onecolumn,
tightenlines,
superscriptaddress,
preprintnumbers,
nofootinbib,
amsmath,amssymb,amsthm,
physrev,
eqsecnum,tikz,
]{revtex4-2}

\usepackage{isomath}
\usepackage{amsmath,amsthm}
\usepackage{amsbsy}
\usepackage{amssymb}
\usepackage{amscd}
\usepackage{amsfonts}
\usepackage{stmaryrd}
\usepackage{siunitx}
\usepackage{euscript}
\usepackage[utf8]{inputenc}
\usepackage[T1]{fontenc}
\usepackage{newtxtext} 
\everymath{\displaystyle}
\usepackage{exscale}
\usepackage{microtype}
\usepackage{booktabs}
\usepackage{algorithm}
\usepackage{algpseudocode}
\usepackage{hyperref}

\usepackage{graphicx}
\usepackage{boxedminipage}
\usepackage{calc}
\usepackage[dvipsnames]{xcolor}
\graphicspath{ {media/} }
\usepackage[caption=false,justification=raggedright]{subfig}

\usepackage{setspace}
\usepackage{enumitem}
\setitemize{noitemsep,topsep=0pt,parsep=0pt,partopsep=0pt}
\setenumerate{noitemsep,topsep=0pt,parsep=0pt,partopsep=0pt}
\setdescription{noitemsep,topsep=0pt,parsep=0pt,partopsep=0pt}

\usepackage[colorinlistoftodos, color=green!40]{todonotes}
\setuptodonotes{inline,caption={}}

\usepackage{soul} 
\usepackage[normalem]{ulem}

\usepackage{orcidlink}
\usepackage{siunitx}
\usepackage[small]{titlesec}

\titlespacing*{\section}{0pt}{12pt plus 4pt minus 2pt}{2pt plus 2pt minus 2pt}
\titlespacing*{\subsection}{0pt}{12pt plus 4pt minus 2pt}{2pt plus 2pt minus 2pt}
\titlespacing*{\subsubsection}{0pt}{12pt plus 4pt minus 2pt}{2pt plus 2pt minus 2pt}
\titlespacing*{\paragraph}{0pt}{12pt plus 4pt minus 2pt}{2pt plus 2pt minus 2pt}

\makeatletter

    \renewcommand*{\p@subsection}{}
    
    \renewcommand*{\p@subsubsection}{}
\makeatother

\usepackage{upgreek}

\newcommand{\half}{\tfrac{1}{2}}

\theoremstyle{definition}

\AtEndEnvironment{definition}{\null\hfill\qedsymbol}

\AtEndEnvironment{remark}{\null\hfill\qedsymbol}

\AtEndEnvironment{example}{\null\hfill\qedsymbol}

\AtEndEnvironment{assumption}{\null\hfill\qedsymbol}

\newcommand{\bfchi}{\mathbold {\chi}}

\newcommand{\bfsigma}{\mathbold {\sigma}}

\newcommand{\bfzero}{\mathbf{0}}

\DeclareMathOperator{\divergence}{div}

\DeclareMathOperator{\trace}{tr}
\DeclareMathOperator{\skw}{skw}

\newcommand{\parderiv}[2]{\frac{\partial #1}{\partial #2}}

\newcommand{\variation}[2]{\updelta_{#2}#1}

\newcommand{\bfm}{{\mathbold m}}
\newcommand{\bfn}{{\mathbold n}}

\newcommand{\bfx}{{\mathbold x}}

\newcommand{\bfC}{{\mathbold C}}

\newcommand{\bfE}{{\mathbold E}}
\newcommand{\bfF}{{\mathbold F}}

\newcommand{\bfI}{{\mathbold I}}

\newcommand{\bfP}{{\mathbold P}}
\newcommand{\bfQ}{{\mathbold Q}}

\newcommand{\bfS}{{\mathbold S}}

\begin{document}


\preprint{To appear in Journal of Applied Mechanics (\href{ https://doi.org/10.1115/1.4071532}{DOI:10.1115/1.4071532})}

\title{Stress Asymmetry in Hard Magnetic Soft Materials}

\author{H. Gökçen Güner}
    \email{hguner@andrew.cmu.edu}
    \affiliation{Department of Civil and Environmental Engineering, Carnegie Mellon University}

\author{Francois Barthelat}
    \affiliation{Department of Mechanical Engineering, University of Colorado}

\author{John D. Clayton}
    \affiliation{Terminal Effects Division, DEVCOM Army Research Laboratory}
    
\author{Carlos Mora-Corral}
    \affiliation{Departamento de Matemáticas, Universidad Autonóma de Madrid}

\author{Noel Walkington}
    \affiliation{Center for Nonlinear Analysis, Department of Mathematical Sciences, Carnegie Mellon University}

\author{Kaushik Dayal \orcidlink{0000-0002-0516-3066}}
    \affiliation{Department of Civil and Environmental Engineering, Carnegie Mellon University}
    \affiliation{Center for Nonlinear Analysis, Department of Mathematical Sciences, Carnegie Mellon University}
    \affiliation{Department of Mechanical Engineering, Carnegie Mellon University}

\date{\today}


\begin{abstract}
    Hard magnetic soft materials -- soft polymers embedded with hard magnetic particles -- are modeled using continuum magnetomechanical formulations in which the deformation and the magnetization field are the primary kinematic variables. 
    A recent question in such formulations is whether the Cauchy stress is symmetric, which is directly related to frame invariance and angular momentum balance. 
    This note discusses energetically equivalent formulations, related by a change of variables between referential and current descriptions of the magnetization, and shows that they generally yield different Cauchy stresses, including a change in their symmetry.
    Specifically, the formulation based on a referential magnetization produces a symmetric Cauchy stress, while that based on a current magnetization generally yields an asymmetric Cauchy stress.
    We highlight that when the internal variable (magnetization field) is at the energy-minimizing equilibrium configuration,  the divergences of these stresses are the same, and both stresses are symmetric.
\end{abstract}

\maketitle
\begin{center}
    \large
    Dedicated to Prof. David J. Steigmann
\end{center}


\section{Introduction}

Hard magnetic soft materials (HMSM), composed of hard magnetic particles that are embedded in a soft polymer matrix, have recently received attention in mechanics due to their potential for applications in robotics and related fields \cite{Kim2018PrintingFerromagnetic,Zhao2019Mechanics, Zhao2022Topology, li2022digital,Wu_2020Multifunctional,Wu2020Multimaterial}.
Homogenized continuum models  -- that is, constitutive magnetomechanical models that do not consider individual particles but only their collective response at larger scales -- use the deformation and magnetization fields as the primary kinematic variables \cite{Zhao2019Mechanics}.

Recent work in HMSM has brought up the issue of frame invariance and the consequent symmetry of the Cauchy stress, e.g., \cite{rahmati2023theory,dorfmann2024hard}.
In this short note, we discuss that equivalent energetic formulations -- which are related by an elementary change of variables -- can generally lead to different Cauchy stresses, including changing whether they are symmetric.

The class of changes of variables that we consider are between referential and current descriptions of the internal variables; the magnetization field, in HMSM.
Here, we emphasize that we are \emph{not} switching between Lagrangian and Eulerian descriptions; all quantities are Lagrangian.
What does change, however, is that the referential and current descriptions transform in different ways under a rigid rotation of the current configuration: the referential description is invariant, whereas the current description is generally not invariant.
A simple example is the temperature gradient: the gradient with respect to referential coordinates is invariant under a rigid rotation of the current configuration, whereas the gradient with respect to current coordinates is not invariant; however, the descriptions are equivalent in that it is possible to transform between them \cite{gurtin1982introduction}.

An elementary example shows the main point of this note.
Consider a function $w(f,n)$, and define $w_0(f,n_0(f,n)) := w(f,n)$ through the change of variables $n_0(f,n)$.
The derivatives are related by the expression:
\begin{equation}
    \begin{Bmatrix}
        \parderiv{w}{f} 
        \\[1em] 
        \parderiv{w}{n}
    \end{Bmatrix}
    =
    \begin{bmatrix}
        1 & \parderiv{n_0}{f} 
        \\[1em]
        0 & \parderiv{n_0}{n}
    \end{bmatrix}
    \begin{Bmatrix}
        \parderiv{w_0}{f} 
        \\[1em] 
        \parderiv{w_0}{n_0}
    \end{Bmatrix}
\end{equation}
It is clear that $\parderiv{w}{f} \neq \parderiv{w_0}{f}$ in general, unless $\parderiv{w}{n}=0$ or equivalently $\parderiv{w_0}{n_0}=0$.
Our central point is that, analogously, the stresses -- defined through the variational derivatives of the energy -- are sensitive to a transformation of the internal variables.

Section \ref{sec:formulation} provides the general framework; Section \ref{sec:examples} discusses simple examples from liquid crystal elastomers; and Section \ref{sec:HMSM} discusses hard magnetic soft matter.

\subsection{Notation.}

\begin{itemize}

    \item We use boldface to denote vectors and tensors.

    \item 
    We use subscripts of $0$ to represent quantities related to the reference configuration; e.g., $\nabla_0$ represents differentiation with respect to the referential position $\bfx_0$, and $\nabla$ to represent differentiation with respect to the deformed position $\bfx$.
    The deformation map is denoted $\chi$, i.e., $\bfx = \chi(\bfx_0)$.
    We use $\bfF = \nabla_0 \bfchi$, $J=\det \bfF$, and $\bfC = \bfF^T \bfF$.
    The standard relation between differentiation with respect to $\bfx_0$ and to $\bfx$ is then $\bfF^{-T} \nabla_0 (\cdot) = \nabla (\cdot)$.

    \item 
    For a 2nd order tensor $\bfsigma$, we use $\skw(\bfsigma):=\half\left(\bfsigma-\bfsigma^T\right)$. 

    \item 
    We use $\delta\bfx$ to denote the variation of a function $\bfx$, and $\variation{A}{\bfx}$ to denote the variational derivative of the functional $A[\bfx]$ with respect to the function $\bfx$.

\end{itemize}

\section{General Framework}
\label{sec:formulation}

Consider an energy $E[\bfx,\bfn]$ with a corresponding energy density $W(\nabla_0\bfx,\bfn)$; that is\footnote{
    Boundary conditions do not play a significant role in this paper, and consequently we assume that they are such that the boundary terms vanish.
}, $E[\bfx,\bfn] = \int_\Omega W(\nabla_0\bfx,\bfn)$.
We use $\bfn$ as a vector internal variable that describes the current value and is written in the Lagrangian sense.

We next introduce $\bfn_0$ as the referential description of $\bfn$.
They are related by the mapping $\bfn = \hat\bfn(\bfF,\bfn_0)$, where $\hat\bfn$ is the function that relates $\bfn_0$ to $\bfn$.
The mapping is simply a change of variables and can be fairly general; for now, we only require that it be invertible and that there exist $\hat\bfn_0$ such that $\bfn_0 = \hat\bfn_0(\bfF,\bfn)$.

We use the mapping between $\bfn$ and $\bfn_0$ to introduce the energy density $W_0(\bfF,\bfn_0)$, defined by:
\begin{equation}
    W_0(\bfF,\bfn_0)
    :=
    W(\bfF,\bfn=\hat\bfn(\bfF,\bfn_0))
\end{equation}
This change of variables leads us to introduce $E_0[\bfx,\bfn_0] = \int_\Omega W_0(\nabla_0\bfx,\bfn_0)$.

\subsection{Implication of Frame Invariance}

We now consider $\bfn$ to be an objective vector.
Then, any rigid rotation $\bfQ$ applied to the current configuration leads to the transformations $\bfF \to \bfQ \bfF$ and $\bfn \to \bfQ\bfn$.
From the usual arguments of frame invariance of $E$, we require:
\begin{equation}
    \label{eqn:invariance-W}
    W(\bfF,\bfn) = W(\bfQ\bfF,\bfQ\bfn), \quad \forall \bfQ \in SO(3)
\end{equation}

Next, we require that $\hat\bfn_0$ satisfies 
$\hat\bfn_0 (\bfQ\bfF,\bfQ\bfn) = \hat\bfn_0 (\bfF,\bfn), \forall \bfQ \in SO(3)$.
Consequently, any rigid rotation $\bfQ$ of the current configuration leaves $\bfn_0$ invariant.
This gives:
\begin{equation}
\label{eqn:invariance-W0}
\begin{split}      
    & W_0(\bfF,\hat\bfn_0(\bfF,\bfn))
    =
    W(\bfF,\bfn)
    =
    W(\bfQ\bfF,\bfQ\bfn)
    =
    W_0(\bfQ\bfF,\hat\bfn_0(\bfQ\bfF,\bfQ\bfn))
    =
    W_0(\bfQ\bfF, \hat\bfn_0(\bfF,\bfn))
    \\
    &\implies W_0(\bfF,\bfn_0)=W_0(\bfQ\bfF,\bfn_0)
\end{split}
\end{equation}

Following the structure of the Noether principle, we apply frame invariance to obtain conditions that relate to angular momentum balance \cite{lubarda2000conservation,knowles1972class,ericksen1976equations,singh2021pseudomomentum,herrmann1981conservation}.
We parameterize the rotation $\bfQ$ by $s$ such that $\bfQ(0) = \bfI, \bfQ'(0) = \bfS$, where $\bfS$ is any skew tensor; explicitly, $\bfQ (s) = e^{s \bfS}$.
We then differentiate \eqref{eqn:invariance-W} with $s$ and evaluate at $s=0$ to get:
\begin{equation}
\label{eqn:symmetry-W}
     \bfS : \left(\parderiv{W}{\bfF}\bfF^T + \parderiv{W}{\bfn}\otimes\bfn\right) = 0
    \implies
    \skw \left(\parderiv{W}{\bfF}\bfF^T + \parderiv{W}{\bfn}\otimes\bfn\right) = \bfzero
\end{equation}
where we have used the arbitrariness of $\bfS$ to conclude that the quantity in parentheses must be symmetric.

A similar procedure applied to \eqref{eqn:invariance-W0} gives:
\begin{equation}
\label{eqn:symmetry-W0}
    \skw \parderiv{W_0}{\bfF}\bfF^T = \bfzero
\end{equation}

\subsection{Symmetry and Asymmetry of the Cauchy Stress}

The minimization of $E_0[\bfx,\bfn_0] = \int_\Omega W_0(\nabla_0\bfx,\bfn_0)$ with respect to its arguments gives the following conditions:
\begin{subequations}
\begin{align}
    \divergence_0 \bfP_0 &= \bfzero, \quad \text{ with } \bfP_0 := \parderiv{W_0}{\bfF}
    \\
    \parderiv{W_0}{\bfn_0} &= \bfzero
\end{align}    
\end{subequations}
where $\bfP_0$ is defined as the Piola-Kirchhoff (P-K) stress associated with $E_0$.
The corresponding Cauchy Stress $\bfsigma_0 = J^{-1} \bfP_0 \bfF^T$ is symmetric from \eqref{eqn:symmetry-W0}.

Similarly, the minimization of $E[\bfx,\bfn] = \int_\Omega W(\nabla_0\bfx,\bfn)$ gives the following conditions:
\begin{subequations}
\begin{align}
    \divergence_0 \bfP & = \bfzero, \quad \text{ with } \bfP := \parderiv{W}{\bfF}
    \\
    \parderiv{W}{\bfn} &= \bfzero
\end{align}    
\end{subequations}

We notice that $\bfP \neq \bfP_0$ in general: while $W(\bfF,\bfn) = W_0(\bfF,\bfn_0)$ for all $\bfF$, we note that $\bfP$ is the partial derivative of $W$ with $\bfn$ fixed and $\bfP_0$ is the partial derivative of $W_0$ with $\bfn_0$ fixed.
Further, from \eqref{eqn:symmetry-W}, $\bfsigma = J^{-1} \bfP \bfF^T$ is not generally symmetric except when the internal variable is at equilibrium, i.e., $\parderiv{W}{\bfn} = \bfzero$.

\subsection{Equivalence of the Referential and Current Descriptions at Internal Variable Equilibrium}

Consider a free energy functional $E[\bfx,\bfn]$ with general dependence on the arguments, including gradients of any order or nonlocal terms.
Here, $\bfn(\bfx_0)$ is a general tensorial quantity that describes the internal variable in the current configuration.
Define the mapping between $\bfn$ and the reference description $\bfn_0$ by $\bfn_0=\hat\bfn_0(\bfF,\bfn)$ and $\bfn=\hat\bfn(\bfF,\bfn_0)$, where the mappings satisfy the requirements discussed above.
We can then define the free energy functional $E_0[\bfx,\bfn_0]$ through:
\begin{equation}
    E_0[\bfx,\bfn_0] := E[\bfx,\bfn=\hat\bfn(\bfF,\bfn_0)]
\end{equation}

Consider the arbitrary admissible variations $\bfx \to \bfx + \delta\bfx, \bfn \to \bfn + \delta\bfn$.
Assuming that the variation of $\bfn_0$ is not arbitrary but is instead given by the mappings between reference and current descriptions, we have the following expression for $\delta\bfn_0$ to leading order:
\begin{equation}
\label{eqn:mapping-deriv}
    \bfn_0 
    \to \bfn_0 + \delta\bfn_0
    = \hat\bfn_0(\bfF + \nabla_0\delta\bfx, \bfn + \delta\bfn)
    \implies
    \delta\bfn_0 = \parderiv{\hat\bfn_0}{\bfF}\nabla_0\delta\bfx + \parderiv{\hat\bfn_0}{\bfn}\delta\bfn
\end{equation}
Since the variation $\delta\bfn_0$ is not arbitrary, but is instead constrained by the reference-current mapping, it follows that the variations $\delta E_0$ and $\delta E$ are equal, giving the relation between the variational derivatives of $\bfE$ and $\bfE_0$:
\begin{equation}
    \variation{E_0}{\bfx}\cdot\delta\bfx
    +
    \variation{E_0}{\bfn_0}\cdot\delta\bfn_0
    =
    \variation{E}{\bfx}\cdot\delta\bfx
    +
    \variation{E}{\bfn}\cdot\delta\bfn    
\end{equation}
where the dot products are interpreted in this section as the Frobenius inner product appropriate for the tensorial order of $\bfn$.

Using \eqref{eqn:mapping-deriv} to replace $\delta\bfn_0$ in the above equation gives:
\begin{equation}
    \left(
        \variation{E_0}{\bfx} - \variation{E}{\bfx}
    \right)\cdot\delta\bfx
    +
    \variation{E_0}{\bfn_0}\cdot\parderiv{\hat\bfn_0}{\bfF}\nabla_0\delta\bfx
    +
    \left(
        \variation{E_0}{\bfn_0} \parderiv{\hat\bfn_0}{\bfn}
        - \variation{E}{\bfn}    
    \right)\cdot\delta\bfn
    =
    0
\end{equation}

Since $\delta\bfn$ is arbitrary, we conclude that $\variation{E_0}{\bfn_0} \parderiv{\hat\bfn_0}{\bfn} = \variation{E}{\bfn}$.
Further, since $\parderiv{\hat\bfn_0}{\bfn}$ is invertible based on the assumptions made on the mapping, it follows that $\variation{E_0}{\bfn_0} = \bfzero \iff \variation{E}{\bfn} = \bfzero$.
In physical terms, the equilibrium conditions with respect to the reference and current descriptions are equivalent.

We next note that the arbitrariness of $\delta\bfx$ does not, in general, imply $\variation{E_0}{\bfx} = \variation{E}{\bfx}$.
However, if we further have $\variation{E_0}{\bfn_0} = \bfzero$, it then follows that $\variation{E_0}{\bfx} = \variation{E}{\bfx}$, where $\variation{E}{\bfx} \equiv \divergence_0 \bfP$ and $\variation{E_0}{\bfx} \equiv \divergence_0 \bfP_0$ represent the resultant forces in the balance of linear momentum.
Therefore, in general, the models predict the same forces in the balance of linear momentum only when the internal variable is allowed to relax, and not necessarily if the internal variable is not relaxed.

\section{Example: Liquid Crystal Elastomers}
\label{sec:examples}

We use Liquid Crystal Elastomers (LCE) as a simple system to observe the implications of Section \ref{sec:formulation}.
We define energies in terms of the current and referential descriptions of a vector internal variable, and compare the resulting stresses.
The energy that we use broadly follows \cite{desimone2009elastic}, specifically the version used in \cite{babaei2021torque}; we also highlight \cite{fried2004free} which has a careful analysis of frame invariance in such models.

The key kinematic variables are the deformation and a vector internal variable $\bfn$ that is denoted the director.
The director describes the current orientation of a molecular-level object, and its effect on the macroscopic mechanics is through a stress-free deformation tensor $\bfF_l = \bfI + \alpha\,\bfn\otimes\bfn$, where $\alpha$ is a material constant.
For simplicity, we do not enforce incompressibility or that $\bfn$ must be a unit vector.

The standard LCE energy, based on the current description $\bfn$, is:
\begin{equation}
    E[\bfx, \bfn]
    =
    \int_{\bfx_0 \in \Omega_0} W(\bfF, \bfn),
    \qquad
    \text{ with } W(\bfF,\bfn)
    = \frac{\mu}{2} \trace (\bfF_e^T\bfF_e) + \half g(|\bfn|^2).
\end{equation}
where the elastic deformation is $\bfF_e = \bfF_l^{-1}\bfF$; $\mu$ is the shear modulus; and $g$ is a function that penalizes $\bfn$ being non-unit.

We define the referential director $\bfn_0(\bfx_0)$ by the mapping $\bfn = \hat\bfn(\bfF,\bfn_0) = \bfF\bfn_0$,
with inverse $\hat\bfn_0(\bfF,\bfn) = \bfF^{-1}\bfn$.
The referential-description energy $E_0$ is then:
\begin{equation}
    E_0[\bfx, \bfn_0]
    =
    \int_{\bfx_0 \in \Omega_0} W_0(\bfF, \bfn_0),
    \qquad
    W_0(\bfF, \bfn_0) := W(\bfF, \bfF\bfn_0)
    = \frac{\mu}{2}\trace(\bfF_e^T\bfF_e\big|_{\bfn=\bfF\bfn_0}) + \half g(|\bfF\bfn_0|^2).
\end{equation}

A rotation $\bfQ$ applied to the current configuration leads to the transformations
$\bfF \to \bfQ\bfF$, $\bfn \to \bfQ\bfn$, and $\bfn_0 \to \bfn_0$.
The invariance of $E_0$ and $E$ under this rotation then provides, respectively:
\begin{subequations}
\label{eqn:LCE-Ericksen}
\begin{align}
    &\skw\left(\parderiv{W_0}{\bfF}\bfF^T\right) = \bfzero
    \\
    &\skw\left(\parderiv{W}{\bfF}\bfF^T + \parderiv{W}{\bfn}\otimes\bfn\right) = \bfzero
\end{align}
\end{subequations}

The equilibrium equations for the director, obtained from $E_0$ and $E$ respectively, are:
\begin{align}
    \variation{E_0}{\bfn_0} = \bfzero &\implies \parderiv{W_0}{\bfn_0} = g'(|\bfF\bfn_0|^2)\bfF^T\bfF\bfn_0 + \mu\parderiv{}{\bfn_0}\trace(\bfF_e^T\bfF_e)\big|_{\bfn=\bfF\bfn_0} = \bfzero
    \\
    \variation{E}{\bfn} = \bfzero &\implies \parderiv{W}{\bfn} = g'(|\bfn|^2)\bfn + \mu\parderiv{}{\bfn}\trace(\bfF_e^T\bfF_e) = \bfzero
\end{align}
Since $\bfF^{-1}$ is invertible, the equilibrium conditions are equivalent: $\variation{E_0}{\bfn_0} = \bfzero \iff \variation{E}{\bfn} = \bfzero$.

The stresses, obtained from $E_0$ and $E$ respectively, are:\begin{align}
    \bfP_0
    &:=
    \parderiv{W_0}{\bfF}\bigg|_{\bfn_0}
    \nonumber\\
    &=
    \mu\left(\bfF_l^{-1}\right)^2\bfF\big|_{\bfn=\bfF\bfn_0}
    +
    \left(
        -\mu\alpha\left(
            \bfF_l^{-1}\bfF_e\bfF_e^T\bfn
            +
            \bfF_e\bfF_e^T\bfF_l^{-1}\bfn
        \right)
        +
        g'(|\bfn|^2)\bfn
    \right)\bigg|_{\bfn=\bfF\bfn_0}\otimes\bfn_0
    \nonumber\\
    &\implies
    \skw\left(\bfsigma_0 := J^{-1}\bfP_0\bfF^T\right) = \bfzero,
    \label{eq:lce-P0}
    \\
    \bfP
    &:=
    \parderiv{W}{\bfF}\bigg|_{\bfn}
    =
    \mu\left(\bfF_l^{-1}\right)^2\bfF
    \nonumber\\
    &\implies
    \skw\left(\bfsigma := J^{-1}\bfP\bfF^T\right)
    =
    -J^{-1}\skw\left(\parderiv{W}{\bfn}\otimes\bfn\right)
    \neq \bfzero.
    \label{eq:lce-P}
\end{align}
where we have used \eqref{eqn:LCE-Ericksen} in evaluating the symmetry of the Cauchy stresses.
From \eqref{eq:lce-P0}--\eqref{eq:lce-P}, the difference in stresses is:
\begin{equation}
    \bfP_0 - \bfP
    =
    \left(\parderiv{W}{\bfn}\otimes\bfn_0\right)
\end{equation}
Hence, at director equilibrium $\parderiv{W}{\bfn} = \bfzero$, the stresses coincide.

\subsection{Effect of Director Gradients}
We augment the LCE energy with a term that penalizes spatial gradients of the director, following the usual model of liquid crystals, e.g. \cite{chandrasekhar1993liquid}.
In terms of $\bfn$, we define:
\begin{equation}
    E[\bfx,\bfn]
    =
    \int_{\bfx_0 \in \Omega_0}
    \left(
        \frac{\mu}{2}\trace(\bfF_e^T\bfF_e)
        +
        \half g(|\bfn|^2)
        +
        \frac{k}{2}|\nabla \bfn|^2
    \right)
\end{equation}
where $\nabla\bfn$ is the spatial gradient $\nabla \bfn$ of $\bfn$:
\begin{equation}
    |\nabla\bfn|^2
    =
    \trace\left(
        \nabla_0\bfn\,(\bfF^T\bfF)^{-1}(\nabla_0\bfn)^T
    \right),
\end{equation}
and we define the energy density:
\begin{equation}
    W(\bfF,\bfn,\nabla_0\bfn)
    =
    \frac{\mu}{2}\trace(\bfF_e^T\bfF_e)
    +
    \half g(|\bfn|^2)
    +
    \frac{k}{2}\trace\left(
        \nabla_0\bfn\,(\bfF^T\bfF)^{-1}(\nabla_0\bfn)^T
    \right).
    \label{eq:frank-W}
\end{equation}

Using the mapping $\bfn = \hat\bfn(\bfF,\bfn_0) = \bfF\bfn_0$, the expression for $E_0$ is:
\begin{equation}
    E_0[\bfx,\bfn_0]
    =
    \int_{\bfx_0\in\Omega_0}
    W_0(\bfF,\nabla_0\bfF,\bfn_0,\nabla_0\bfn_0),
\end{equation}
with
\begin{equation}
\begin{split}
    W_0(\bfF,\nabla_0\bfF,\bfn_0,\nabla_0\bfn_0)
    :=
    &
    \frac{\mu}{2}\trace\left(
        \left.\bfF_e^T\bfF_e\right|_{\bfn=\bfF\bfn_0}
    \right)
    +
    \half g\left(|\bfF\bfn_0|^2\right)
    \\
    &
    +\frac{k}{2}\trace\left(
        \Bigl((\nabla_0\bfF)\,\bfn_0+\bfF\,\nabla_0\bfn_0\Bigr)
        (\bfF^T\bfF)^{-1}
        \Bigl((\nabla_0\bfF)\,\bfn_0+\bfF\,\nabla_0\bfn_0\Bigr)^T
    \right)
\end{split}
\label{eq:frank-W0}
\end{equation}

The equilibrium equations for the director, obtained from $E_0$ and $E$ respectively, are
\begin{align}
    \variation{E_0}{\bfn_0}=\bfzero
    &\implies
    \parderiv{W_0}{\bfn_0}
    -
    \divergence_0\parderiv{W_0}{\nabla_0\bfn_0}
    =
    \bfzero,
    \label{eq:frank-EL-ref}
    \\
    \variation{E}{\bfn}=\bfzero
    &\implies
    \parderiv{W}{\bfn}
    -
    \divergence_0\left(
        k\,\nabla_0\bfn\,(\bfF^T\bfF)^{-1}
    \right)
    =
    \bfzero.
    \label{eq:frank-EL-cur}
\end{align}
Since $\bfF$ is invertible, these two director equilibrium conditions are equivalent: $\variation{E_0}{\bfn_0}=\bfzero \iff \variation{E}{\bfn}=\bfzero$, where \eqref{eq:frank-EL-ref} is related by $\bfF^T$ to \eqref{eq:frank-EL-cur}.

The stresses, obtained from $E_0$ and $E$ respectively, are
\begin{align}
    \bfP
    &:=
    \parderiv{W}{\bfF}\bigg|_{\bfn,\,\nabla_0\bfn}
    =
    \mu\left(\bfF_l^{-1}\right)^2\bfF
    -
    k\,\bfF^{-T}(\nabla_0\bfn)^T\nabla_0\bfn\,(\bfF^T\bfF)^{-1},
    \label{eq:frank-P}
    \\
    \bfP_0
    &:=
    \parderiv{W_0}{\bfF}
    -
    \divergence_0 \parderiv{W_0}{\nabla_0 \bfF}
    \nonumber\\
    &=
    \mu\left(\bfF_l^{-1}\right)^2\bfF\big|_{\bfn=\bfF\bfn_0}
    -
    k\,\bfF^{-T}(\nabla_0\bfn)^T\nabla_0\bfn\,(\bfF^T\bfF)^{-1}\bigg|_{\bfn=\bfF\bfn_0}
    \nonumber\\
    &\quad
    +
    \left(
        -\mu\alpha\Bigl(
            \bfF_l^{-1}\bfF_e\bfF_e^T\bfn
            +
            \bfF_e\bfF_e^T\bfF_l^{-1}\bfn
        \Bigr)
        +
        g'(|\bfn|^2)\bfn
        -
        \divergence_0\left(
            k\,\nabla_0\bfn\,(\bfF^T\bfF)^{-1}
        \right)
    \right)\bigg|_{\bfn=\bfF\bfn_0}\otimes\bfn_0.
    \label{eq:frank-P0}
\end{align}
A direct calculation gives:
\begin{equation}
    \bfP_0
    =
    \bfP
    +
    \left(
        \parderiv{W}{\bfn}
        -
        \divergence_0\left(
            k\,\nabla_0\bfn\,(\bfF^T\bfF)^{-1}
        \right)
    \right)\otimes\bfn_0.
    \label{eq:frank-P0-compare}
\end{equation}
When the director is at the equilibrium configuration \eqref{eq:frank-EL-cur}, we have $\bfP_0=\bfP$.

\section{Hard Magnetic Soft Materials}
\label{sec:HMSM}

The starting point is the free energy, where the local part is posed in the reference configuration and the electromagnetic (EM) terms are naturally posed in the current configuration following, e.g.,  \cite{james1993theory,desimone2002constrained,steigmann2004equilibrium,steigmann2009formulation,liu2014energy,grasinger2021architected,khandagale2025nonlocal}:
\begin{equation}
\label{eqn:HMSM-energy-1}
\begin{split}
    \tilde{E}[\bfx,\tilde\bfm]
    =
    &
    \int_{\bfx_0\in\Omega_0} \tilde W(\bfF,\tilde\bfm(\chi(\bfx_0))) + \frac{\mu_0}{2} \int_{\bfx\in\mathbb{R}^3} |\nabla\phi|^2
    \\
    & \qquad
    \text{ with $\phi$ the solution of } \divergence\nabla\phi = \divergence\tilde{\bfm} \text{ on } \mathbb{R}^3   
\end{split}
\end{equation}
where $\tilde\bfm(\bfx)$ is the current magnetization as a function of the current configuration.
This form has the structure of a constrained minimization problem as opposed to a saddle-point problem \cite{james1990frustration}, which is important for several reasons including providing robust numerical methods, e.g. \cite{yang2011completely}.

While the EM contribution to the energy originates in the current description, it is convenient to apply a change of variables such that the integration and differentiation operations are performed with respect to $\bfx_0$.
Following\footnote{See Section 4.1.1 in \cite{marshall2014atomistic} for a discussion of different pullback mappings and their interpretation as differential geometric forms; \cite{sky2025cosserat} motivate their pullbacks based on similar ideas.} \cite{dorfmann2005nonlinear, castaneda2012finite, liu2014energy}, we define the reference magnetization $\bfm_0(\bfx_0)$, where $\tilde\bfm(\chi(\bfx_0)) = J(\bfx_0)^{-1}\bfF(\bfx_0) \bfm_0 (\bfx_0)$, which has the form:
\begin{equation}
\begin{split}
    E_0[\bfx,\bfm_0]
    =
    &
    \int_{\bfx_0\in\Omega_0} W_0(\bfF,\bfm_0) + \frac{\mu_0}{2} \int_{\bfx_0\in\mathbb{R}^3} J \nabla_0\phi\cdot\bfC^{-1}\nabla_0\phi
    \\
    & \qquad
    \text{ with $\phi$ the solution of } \divergence_0 J \bfC^{-1} \nabla_0\phi = \divergence_0\bfm_0 \text{ on } \mathbb{R}^3   
\end{split}
\end{equation}

We next rewrite $E_0$ in terms of the current magnetization written as a function of $\bfx_0$, using the mapping $\bfm(\bfx_0) = J(\bfx_0)^{-1}\bfF(\bfx_0) \bfm_0 (\bfx_0)$, to get:
\begin{equation}
\begin{split}
    E[\bfx,\bfm]
    =
    &
    \int_{\bfx_0\in\Omega_0} W(\bfF,\bfm) + \frac{\mu_0}{2} \int_{\bfx_0\in\mathbb{R}^3} J \nabla_0\psi\cdot\bfC^{-1}\nabla_0\psi
    \\
    & \qquad
    \text{ with $\psi$ the solution of } \divergence_0 J \bfC^{-1} \nabla_0\psi = \divergence_0 J \bfF^{-1} \bfm \text{ on } \mathbb{R}^3   
\end{split}
\end{equation}

\subsection{Implications of Frame Invariance}

A rotation $\bfQ$ applied to the current configuration leads to the transformations $\bfF \to \bfQ \bfF$, $\bfm \to \bfQ\bfm$, and $\bfm_0 \to \bfm_0$.
This transformation leaves invariant the EM fields $\nabla_0 \phi$ and $\nabla_0\psi$.
Consequently, the EM energy terms in both $E_0$ and $E$ are invariant.
The invariance of $E_0$ and $E$ then provide, respectively:
\begin{subequations}
\label{eqn:HMSM-Ericksen}
\begin{align}
    & \skw \left( \parderiv{W_0}{\bfF}\bfF^T \right)=\bfzero 
    \\
    & \skw \left( \parderiv{W}{\bfF}\bfF^T + \parderiv{W}{\bfm}\otimes\bfm \right)=\bfzero
\end{align}
\end{subequations}

\subsection{Stresses and Equilibrium Conditions for the Magnetization}

The variational derivatives of $E_0$ and $E$ are computed using standard techniques, e.g., following \cite{james2002configurational,liu2014energy,marshall2014atomistic}.
The equilibrium equations for the magnetization, obtained from $E_0$ and $E$ respectively, are:
\begin{subequations}
\label{eqn:HMSM-equilib}
\begin{align}
    \variation{E_0}{\bfm_0} = 0 & \implies \parderiv{W_0}{\bfm_0} + \mu_0 \nabla_0 \phi = \bfzero
    \\
    \variation{E}{\bfm} = 0 & \implies \parderiv{W}{\bfm} + \mu_0 J \bfF^{-T} \nabla_0 \psi = \bfzero
\end{align}
\end{subequations}
The stresses, obtained from $E_0$ and $E$ respectively, are:
\begin{subequations}
\label{eqn:HMSM-stress}
\begin{align}
    \bfP_0 & 
    :=
    \parderiv{W_0}{\bfF} - \frac{\mu_0}{2} J |\bfF^{-T} \nabla_0\phi|^2 \bfF^{-T} + \mu_0 J \bfF^{-T}\nabla_0\phi \otimes \bfC^{-1}\nabla_0\phi
    \implies
    \skw \left(\bfsigma_0 := J^{-1} \bfP_0 \bfF^T\right) = \bfzero    
    \\
    \begin{split}
    \bfP & 
    := 
    \parderiv{W}{\bfF} 
    - \frac{\mu_0}{2} J |\bfF^{-T} \nabla_0\psi|^2 \bfF^{-T} 
    + \mu_0 J \bfF^{-T}\nabla_0\psi \otimes \bfC^{-1}\nabla_0\psi 
    + \mu_0 J \left(\bfF^{-T}\nabla_0\psi\cdot\bfm \right)\bfF^{-T}
    - \mu_0 J \bfF^{-T}\nabla_0\psi \otimes \bfF^{-1}\bfm 
    \\
    & \implies
    \skw\left(\bfsigma := J^{-1} \bfP \bfF^T\right)  
    =
    -\skw\left(\left(J^{-1}\parderiv{W}{\bfm} + \mu_0 \bfF^{-T}\nabla_0\psi\right)\otimes\bfm\right)
    \neq \bfzero
    \end{split}
\end{align}
\end{subequations}
where we have used the invariance relations \eqref{eqn:HMSM-Ericksen} in computing the skew parts of the Cauchy stresses.
We notice that there are additional terms in $\bfP$ compared to $\bfP_0$.
These arise due to the mapping $\bfm = J^{-1} \bfF \bfm_0$; other choices of mappings lead to different expressions of $\bfP$, e.g., \cite{rahmati2023theory,katusele2025exploiting,katusele2025soft}.

An important implication from \eqref{eqn:HMSM-stress} is that $\bfsigma_0$ is symmetric while $\bfsigma$ is not.
However, we highlight that $\skw\bfsigma$ vanishes precisely at magnetic equilibrium \eqref{eqn:HMSM-equilib}.


\section{Concluding Remarks}

We have shown that energetically equivalent continuum models with vectorial internal variables can give rise to different stresses, even permitting a change in symmetry of the Cauchy stress.
In the context of hard magnetic soft matter, we find that the magnetomechanical Maxwell stresses are symmetric in formulations based on referential descriptions of the magnetization but are not necessarily symmetric in formulations based on current descriptions of the magnetization.
The key issue is that the stresses are obtained from the energy as partial derivatives with respect to the deformation while keeping the internal variable fixed; since the internal variables are coupled to the deformation, a transformation of the internal variable naturally leads to a different value for the partial derivatives.
We note that we have not included external work terms in our discussion, but it is a simple extension that would not change our conclusions.

However, we also find that key aspects of the various formulations coincide when the internal variable is at equilibrium, i.e., the derivative of the energy with respect to the internal variable vanishes.
Under this condition, while the stress is not necessarily the same, the resultant divergence of the stress is the same and, in the case of hard magnetic soft matter, the stresses are symmetric.
However, when the internal variable is evolving away from equilibrium -- e.g., through Landau-Lifshitz-Gilbert dynamics for magnetism \cite{lakshmanan2011fascinating}, or through Ericksen-Leslie dynamics for liquid crystals \cite{chandrasekhar1993liquid} -- the stresses and their resultants can generally be different.

\section*{}

\paragraph*{Acknowledgments.}

We thank Richard D. James, Anthony Rollett, and Pradeep Sharma for useful discussions; 
and ARO (MURI W911NF-24-2-0184) for financial support.

\newcommand{\etalchar}[1]{$^{#1}$}

\end{document}